# Numerical Investigation of Laser-Assisted Nanoimprinting on a Copper Substrate from a Perspective of Heat Transfer Analysis

Chun-Ping Jen

Department of Mechanical Engineering, National Chung Cheng University,
No. 168, University Rd., Min-Hsiung, Chia Yi, 62102, Taiwan, R.O.C.

*Abstract-* The technique of laser-assisted nanoimprinting lithography (LAN) has been proposed to utilize an excimer laser to irradiate through a quartz mold and melts a thin polymer film on the substrate for micro- to nano-scaled fabrications. In the present study, the novel concept of that copper was adopted as the substrate instead of silicon, which is conventionally used, was proposed. The micro/nano structures on the copper substrate could be fabricated by chemical/electrochemical etching or electroforming; following by the patterns have been transferred onto the substrate using LAN process. Alternatives of the substrate materials could lead versatile applications in micro/nano-fabrication. To demonstrate the feasibility of this concept numerically, this study introduced optical multiple reflection theory to perform both analytical and numerical modeling during the process and to predict the thermal response theoretically.

## I. INTRODUCTION

Nano-imprinting lithography (NIL) has been developed over a decade [1-5] and is now a promising method for nano-patterning and nano-fabrication. The basic concept of conventional nano-imprinting includes a mold, an etching resist layer and a sample substrate. The mold has some nano-scale features on the surface fabricated by either E-beam lithography or focus ion bean techniques. The thermo-plastic polymer such as poly-methylmethacrylate (PMMA) is usually used as the resist layer. By heating until above its glass transition temperature (Tg), the mold can impinge into the resist layer and form a pattern. Following by reaction ion etching, the nano-pattern is transformed to the resist layer and the substrate. It is sequent followed by standard lithography processes to achieve nano-structures on the substrate surface. However, the heating mechanism of conventional nanoimprinting is slow and usually causes misalignment due to the different thermal expansion between the mold and the substrate. Recently, laser-assisted nanoimprinting lithography (LAN) method illustrated in Figure 1a was proposed [6]. This new imprinting method shares the same concepts discussed above and combines the advantages of the traditional nano-imprinting and laser-assisted direct imprint [4]. The LAN process depicted schematically in Figure 1a utilizes a laser pulse to irradiate through a quartz mold and melt a thin polymer film (~200 nm) on the substrate. Upon the thin polymer film is melted, the pre-loaded mold is imprinted into the polymer resist layer. This LAN method has several obvious advantages over the thermal-based nano-imprinting technique in terms of shortening the processing time and reducing the heating of the substrate and in this way the misalignment due to the thermal expansion mismatch of the mold and substrate can be avoided [6]. This method shows a great potential for future nano-patterning and fabrication of nano-structures. In the present study, the novel concept of that copper was adopted as the substrate instead of silicon, which is conventionally used, was proposed. The micro/nano structures on the copper substrate could be fabricated using chemical/electrochemical etching or electroforming; following by the patterns have been transferred onto the substrate using LAN process. Alternatives of the substrate materials could lead versatile applications in micro/nano-fabrication. To demonstrate the feasibility of this concept numerically, this study introduced optical multiple reflection theory to perform both analytical and numerical modeling during the process and to predict the thermal response theoretically.

## II. THEORETICAL MODELING

During the heating process of the substrate by the pulsed laser, one-dimensional transient heat-diffusion equation can be employed herein. The physical domain for the heat-transfer system considered here is illustrated in Figure 1b. Assuming that the laser spot is larger than the mold and all materials involved are homogeneous; therefore this system can be simplified as one-dimensional. The source term of heat-generation in the substrate originating from the energy absorption of a laser pulse can be described as:

$$S(x,t) = A_S \beta \exp(-\beta x) I(t) \qquad (1)$$

where x and t are the spatial variable and time, respectively. $A_S$ is the absorption of the incident light, $\beta$ is the optical absorption coefficient which is related to the imaginary part of refractive index, $\kappa$, and the laser wavelength, $\lambda$, by $\beta = \dfrac{4\pi\kappa}{\lambda}$ [8]. I(t) is the power density function of the incident laser pulse (power per unit area, W/m$^2$). A Gaussian-shaped laser pulse is assumed in the present study and the power density function can be expressed as:

$$I(t) = \frac{I_{\max}}{\sqrt{2\pi}\sigma} \exp\left(\frac{-(t-\tau_{\max})^2}{2\sigma^2}\right) \qquad (2)$$

where $I_{max}$ is the maximum power density of the laser pulse at $t = \tau_{max}$ and $\sigma$ the standard deviation of the function and can be determined by the full-width at half-maximum (FWHM, $t_0$) of the pulse, $t_0 = 2\sqrt{2\ln 2}\sigma$. $\tau_{max}$ is chosen as the triple of the standard deviation ($\sigma$) and the irradiating time is the double of $\tau_{max}$. For the LAN process, the incident laser propagates through four materials: air (vacuum); quartz mold (fused silica); thin film (polymer) and copper substrate. The optical absorbance in the polymer thin film should be calculated considering optical multiple interferences. Hence, the heat generation in the thin film and the intensity of the excimer laser passing into the substrate can be obtained. The calculation matrix of electric vector based on the fundamental equations of optical characteristic matrix method [9] can be expressed as followings:

$$\begin{bmatrix} E_{m-1}^+ \\ E_{m-1}^- \end{bmatrix} = \frac{1}{t_m} \begin{bmatrix} e^{i\delta_{m-1}} & r_m e^{i\delta_{m-1}} \\ r_m e^{-i\delta_{m-1}} & e^{-i\delta_{m-1}} \end{bmatrix} \begin{bmatrix} E_m^+ \\ E_m^- \end{bmatrix} \quad (3)$$

$$t_m = \frac{2n_{m-1}}{n_{m-1} + n_m} \quad (4)$$

$$r_m = \frac{n_{m-1} - n_m}{n_{m-1} + n_m} \quad (5)$$

$$\delta_m = \frac{2\pi n_m d_m}{\lambda} \quad (6)$$

where $E_m^+$ and $E_m^-$ are the forward and backward electric waves at the interface of the $m^{th}$ layer, respectively. $n_m$ and $d_m$ are the complex refractive index and the thickness of the $m^{th}$ layer, respectively. The calculation is for normal incident radiation and assumes the materials are optically homogeneous. After solving the above matrix, the electric vectors in each layer are obtained. Therefore, the energy absorbance $A_m$ in the $m^{th}$ layer can be calculated by the following equation:

$$A_m = \left( \text{Re}(n_{m-1}) \cdot |E_{m-1}^+|^2 - \text{Re}(n_m) \cdot |E_m^+|^2 \right) + \left( \text{Re}(n_m) \cdot |E_m^-|^2 - \text{Re}(n_{m-1}) \cdot |E_{m-1}^-|^2 \right) \quad (7)$$

In the present investigation, the transmission at air(vacuum)-quartz mold interface is calculated using simple definition of the optical transmissivity between two materials, i.e. $T_a = \frac{4n_0 n_{Quartz}}{(n_0 + n_{Quartz})^2}$. The incident laser is then considered as propagating through quartz mold, polymer resist layer and substrate. The energy absorbed in the polymer resist layer and the substrate is calculated by the model of the optical multiple interference discussed above. The energy absorbance is assumed as uniform distributed in the polymer resist layer. Furthermore, the incident energy entering the substrate is absorbed and distributed as the heat source formulated in Eq. (1).

First of all, convective and radiative heat losses at the surface of the substrate and heat absorption by the quartz mold are neglected and the properties of all materials are assumed to be constant (temperature-independent). The fused quartz mold and the polymer resist layer are treated as highly transparent to laser wavelength without energy absorption. Furthermore, the heat conduction of the quartz and polymer are so small that the surface of substrate can be treated as a thermal-insulated boundary. As a result of these assumptions, the mold and polymer resist layer could be reasonably excluded from the heat transfer system and the problem could be simplified as a semi-infinite slab ($0 < x < \infty$) as shown in Figure 1c. The superposing effects of a large number of instantaneous sources (one-shot) are adopted to obtain the time- and spatial-dependent temperature profile analytically. First, consider the x-direction through an infinite domain with constant properties. The equation for time-dependent heat conduction in the x-direction is expressed as following:

$$\frac{\partial \theta}{\partial t} = \alpha \frac{\partial^2 \theta}{\partial x^2} \quad (8)$$

where $\theta(x,t) = T(x,t) - T_i$ is the excess temperature relative to the initial (far-field) temperature of the domain ($T_i$). $\alpha = \frac{k}{\rho C}$ is the thermal diffusivity of the material. $\rho$ denotes the density, $C$ represents the specific heat, and $k$ is the material's thermal conductivity. The excess temperature distribution of the infinite domain in which an instantaneous plane source, $q''$ (J/m$^2$), is released once at t=0 is [10]:

$$\theta(x,t) = \frac{q''}{2\rho C(\pi \alpha t)^{1/2}} \exp\left(-\frac{x^2}{4\alpha t}\right) \quad (9)$$

The initial and boundary conditions in the modeling domain can be expressed as:

$T(x,t=0) = T_i$; $T(x=\infty, t) = T_i$;

$$\left.\frac{\partial T}{\partial x}\right|_{x=0} = 0 \quad (10)$$

where $T_i$ is the initial temperature of the substrate. Since the gradient of temperature at x=0 equals to zero, the calculated domain can be extend to an infinite medium ($-\infty < x < \infty$) which is symmetric at x=0. According to the concept of the superposition procedure, the medium which is heated by the heat-generation sources, $S(x,t)$, the time-dependent temperature profile is solved analytically and expressed as:

$$\theta(x,t) = \int_0^{t_p} \int_{-\infty}^{\infty} \frac{S(\xi,\tau)}{2\rho C[\pi\alpha(t-\tau)]^{1/2}} \exp\left[-\frac{(x-\xi)^2}{4\alpha(t-\tau)}\right] d\xi\, d\tau \quad (11)$$

where $t_p$ is the irradiating time of a laser pulse. Thus, the temperature inside the substrate is rewritten as following:

$$T(x,t) = \int_0^{t_p} \int_{-\infty}^{\infty} \frac{S(\xi,\tau)}{2\rho C[\pi\alpha(t-\tau)]^{1/2}} \exp\left[-\frac{(x-\xi)^2}{4\alpha(t-\tau)}\right] d\xi\, d\tau + T_i$$

for $x \geq 0$, $t \geq 0$ (12)

The calculated domain has been extended to an infinite medium; hence, the virtual heat sources have to impose in the negative region of the medium to conserve the total energy absorbed in the substrate. The source term S(x,t) should be written as:

$$S(\xi,\tau) = A_S \beta \exp(-\beta|\xi|) I(\tau) \text{ for } -\infty < \xi < \infty \quad (13)$$

The analytical solutions of the transient temperature distribution in Eq. (12) can be integrated using 104-point Gauss-Legendre quadrature.

To solve the transient temperature distribution of the entire system, i.e. quartz/polymer/substrate, the governing equation of energy conservation coupled with the initial and boundary conditions as well as the source term could be numerically solved by a fully implicit finite difference method [11]. This method employs a two point backward difference approximation in time (Euler implicit scheme), a central difference approximation for the second derivative in the spatial coordinate, and an upwind difference approximation for the first derivative in the spatial coordinate. Since the thermal-affected region in the entire system is confined to the region nearby the surface of the substrate, the computational domains of quartz mold and substrate are considered as 200 μm.

### III. RESULTS

Table 1 listed the thermal properties and optical parameters of materials used in this study. Excimer lasers with three different wavelengths (ArF, 193 nm; KrF, 248nm; XeCl, 308nm) are considered as pulsed laser sources herein. Figure 2 shows the surface temperature of copper and silicon substrate irradiated by XeCl excimer laser with 20 ns pulse duration. The temperature increases as the laser fluence increases. The temperature at the surface of copper substrate is much lower than that at silicon substrate due to the smaller absorption coefficient of copper. However, the literature suggests the laser fluence (XeCl, 20 ns, single pulse) irradiating on silicon substrate should be at least 0.35 J/cm$^2$ (Xia et al. 2003). The peak temperature at silicon surface should be around 700 °C when the laser fluence is 0.35 J/cm$^2$, as shown in Fig. 2b. Therefore, the required laser fluence irradiates on the copper substrate could be adopted as 0.6 J/cm$^2$ according to the results in Fig. 2a, which could be the suitable condition for the LAN process. The surface temperature histories of copper substrate irradiated by XeCl excimer laser with fluence of 0.6 J/cm$^2$ for different pulse duration are depicted in Fig. 3. Both the results obtained from the analytical and the numerical approaches are shown in this figure. The numerical approach considers the existence of fused silica mold and polymer resist and the results exhibit the analytical results are almost the same with the numerical results, which means the existence of mold and polymer resist could be ignored, i.e., the assumption of the thermal-insulated boundary mentioned above is reasonable. The increase of the pulse duration decreases the maximum temperature at the surface of the substrate (several hundreds of Celsius degrees). The temperature profiles with time at the different locations inside the substrate with the irradiation of 0.6 J/cm$^2$ of a XeCl excimer laser (pulse duration is 20 ns) are depicted in Fig. 4. The temperature drops dramatically as shown in this figure, for instance, the maximum temperature is about 760 °C at the surface and drops less than 150 °C at 5 μm in depth. The temperature affected region is limited within the extremely shallow depth in the substrate in the same manner as employing silicon as the substrate conventionally. To investigate the temperature of the polymer resist layer, the numerical results at the center of the polymer resist layer and the substrate's surface with the irradiation of 0.6 J/cm$^2$ of a XeCl excimer laser for different pulse durations are illustrated in Fig. 5. The results indicate the maximum temperature at the center of the polymer resist layer decreases slightly and the time which the maximum temperature occurs delays tens of nanoseconds while the pulse duration increases. The temperature in the polymer resist layer exceeds the polymer glass transition temperature (about 100 °C) for all cases. The process time of LAN will increase several tens of nanoseconds while increasing the pulse duration of the excimer laser; meanwhile, the maximum temperature at the silicon's surface could be lowered for several hundreds of degrees. The results successfully demonstrate the feasibility that copper could be adopted as the substrate instead of silicon, which is conventionally used. The heating processes under different laser sources, such as ArF (193 nm), KrF (248 nm) and XeCl (308 nm) are discussed in the following figure. Figure 6 illustrates the temperature profiles with time at the center of the polymer resist layer and the substrate's surface for different laser sources when the laser fluence is 0.6 J/cm$^2$ and the pulse duration is 30 ns. The results in Fig. 6 reveal the temperatures at copper surface and the polymer for these cases are almost the same because copper is insensitive to the three wavelengths. Based on the optical simulation above-mentioned, the energy absorbances in the copper substrate are 71.42%, 73.09% and 73.13% for ArF, KrF and XeCl laser sources, respectively. As shown in the insert in Fig. 6, the temperature at the copper's surface irradiated by ArF is slightly lower than that by KrF and XeCl. However, the energy absorbances in the silicon substrate are 39.5%, 42.1% and 53.3% for ArF, KrF and XeCl laser sources, respectively. The copper substrate absorbed much higher energy from the excimer laser pulse (around 20% more) than the silicon substrate, nevertheless, the temperate at the copper substrate is much lower than that at silicon substrate. This is owing to larger thermal conductivity of copper, which is more than two-fold of that of silicon.

### IV CONCLUSIONS

This work has considered optical multiple reflection theory to perform both analytical modeling and numerical simulation of laser induced heating during LAN process on a copper substrate and to predict the thermal response of the polymer resist and substrate theoretically. The analytical modeling ignored the existence of the fused silica mold and polymer resist; however, as shown in the results, the analytical solutions were well agreed with the numerical simulation, which considered the existence of the fused silica mold and polymer resist. The temperature in the polymer resist layer exceeds the polymer glass transition temperature (about 100 °C) while the copper substrate irradiated by the excimer laser with the fluence of 0.6 J/cm$^2$ presented in the present study.

The results successfully demonstrate the feasibility that copper could be adopted as the substrate instead of silicon, which is conventionally used. The results revealed the increase of the pulse duration could decrease the maximum temperature at the surface of the substrate dramatically and increase only several tens of nanoseconds sequentially. The temperatures at copper surface and the polymer for these cases are almost the same because copper is insensitive to the laser sources, which are ArF, KrF and XeCl. The copper substrate absorbed much higher energy from the excimer laser pulse (around 20% more) than the silicon substrate, nevertheless, the temperate at the copper substrate is much lower than that at silicon substrate owing to larger thermal conductivity of copper.

Table 1 Thermal properties and optical parameters of materials used herein.

| Substrate | Si | Copper[b] | PMMA | Fused Silica[f] |
|---|---|---|---|---|
| Density (kg/m$^3$) | 2300[a] | 8940 | 1170[c] | 2201 |
| Thermal conductivity (W/mK) | 160[a] | 352 | 0.16[d] | 1.30 |
| Heat capacity (J/kgK) | 707.71[a] | 384.9 | 1380[c] | 787.52 |
| Refractive index, ArF / Absorption coefficient (nm$^{-1}$) | (0.872, 2.757)[b] / 0.1795[b] | (0.972, 1.403) / 0.0914 | 1.492[e] | (1.560841, 0.0) / 0.0 |
| Refractive index, KrF / Absorption coefficient (nm$^{-1}$) | (1.570, 3.565)[b] / 0.1806[b] | (1.470, 1.780) / 0.0902 | 1.492[e] | (1.508601, 0.0) / 0.0 |
| Refractive index, XeCl / Absorption coefficient (nm$^{-1}$) | (5.013, 3.689)[b] / 0.1505[b] | (1.350, 1.710) / 0.0698 | 1.492[e] | (1.485663, 0.0) / 0.0 |

[a]Campbell 2001; [b]Lide, 2003; [c]Van Krevelen 1976; [d]Chu et al. 2001; [e]http://www.io.tudelft.nl/research/dfs/idemat/Onl_db/Id123p.htm; [f]http://www.corning.com/.

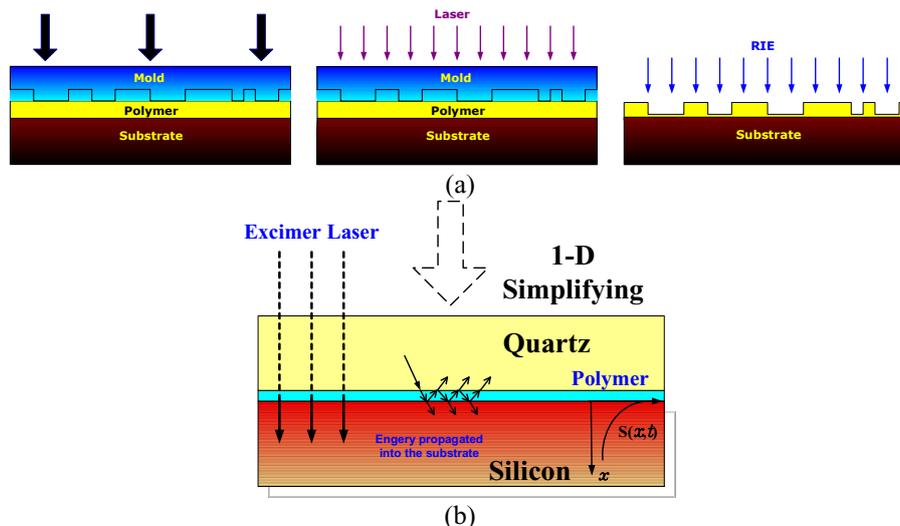

Figure 1: Schematic diagram of nanoimprinting processes: (a) Laser-assisted nanoimprinting lithography (LAN); and (b) illustration of the LAN imprinting process and one-dimensional simplification.

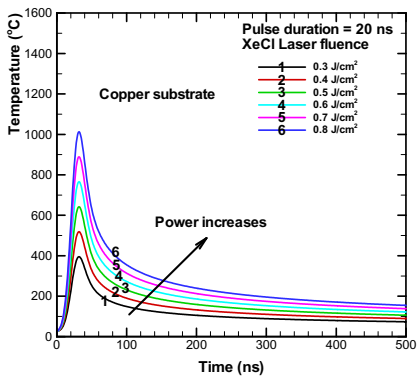

(a)

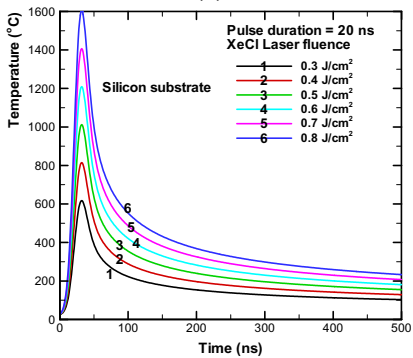

(b)

Figure 2: The surface temperature of (a) copper and (b) silicon substrate irradiated by XeCl excimer laser with 20 ns pulse duration under different laser fluences.

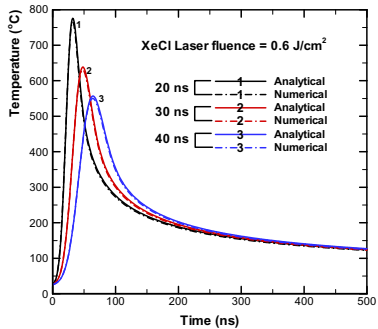

Figure 3: The surface temperature histories of copper substrate irradiated by XeCl excimer laser with fluence of 0.6 J/cm$^2$ for different pulse duration obtained from analytical approximation and numerical simulation which considered the existence of fused silica mold and polymer resist layer.

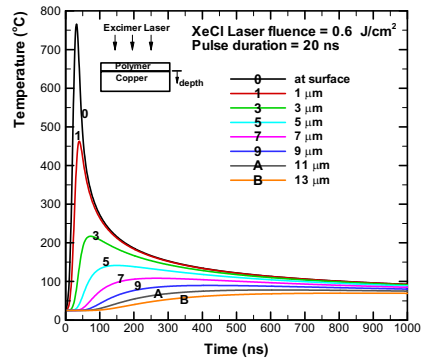

Figure 4: The temperature profiles with time at the different locations inside the substrate with the irradiation of 0.6 J/cm$^2$ of a XeCl excimer laser, and the pulse duration is 20 ns.

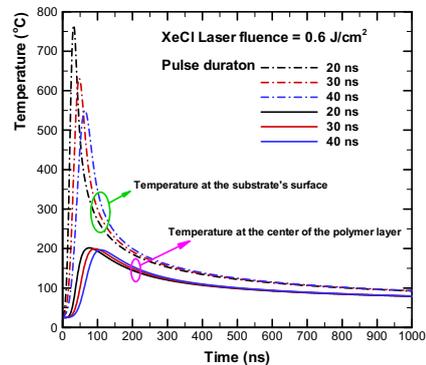

Figure 5: The temperature profiles with time at the center of the polymer resist layer and the substrate's surface when the XeCl laser fluence is 0.6 J/cm$^2$ under different pulse durations.

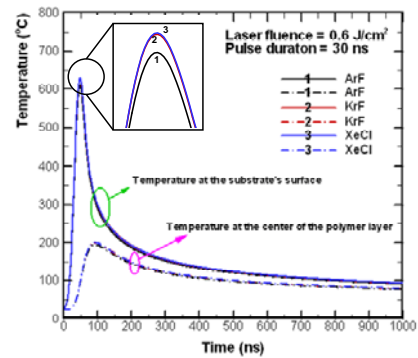

Figure 6: The temperature profiles with time at the center of the polymer resist layer and the substrate's surface for different laser sources when the laser fluence is 0.6 J/cm$^2$ and the pulse duration is 30 ns.